\documentstyle[aps,prl,epsf,floats,amsmath,amssymb]{revtex}

\begin{document}
\draft

\twocolumn[\hsize\textwidth\columnwidth\hsize\csname@twocolumnfalse%
\endcsname

\title  {Critical behaviour of the $1d$ annihilation fission process $2A\to\text{\O}$,
$2A\to 3A$}

\author{G\'eza \'Odor}
\address{Research Institute for Technical Physics and Materials Science, \\
H-1525 Budapest, P.O.Box 49, Hungary}    
\maketitle

\begin{abstract}
Numerical simulations and cluster mean-field 
approximations with coherent anomaly extrapolation
show that the critical line of the $1d$ annihilation fission 
process is separated into two regions. 
In both the small and high diffusion
cases the critical behavior is different from the well known
universality classes of non-equilibrium phase transitions
to absorbing states. The high diffusion region seems to
be well described by the cyclically coupled directed 
percolation and annihilating random walk. Spreading 
exponents show non-universal behavior.
\end{abstract}

]

Non-equilibrium phase transition may take place even in
one dimensional systems. However the ordered phase lacks 
fluctuations that could destroy the state. If the system 
has evolved into that state it will be trapped there.
We call this transition to an absorbing state.
First order transition among these is very rare, 
there have been only a few system found that exhibit a 
discontinuous jump from the active to the absorbing phase 
\cite{TCM,MeOd95,OSz96,MeOd98,Oerding99}.

Continuous phase transitions have been found to belong to
a few universality classes, the most robust of them is the
directed percolation (DP) class. According to the 
hypothesis of \cite{Jan81,Gras82} all continuous phase 
transitions to a single absorbing state in homogeneous 
systems with short ranged interactions belong to this class
provided there is no additional symmetry and quenched 
randomness present. The most prominent system exhibiting
phase transition of this class is the 
branching and annihilating random walk with one offspring
(BARW). Furthermore systems with infinitely
many absorbing states were also found to belong to this
class \cite{PCP,Mendes-Marques}. 

An important exception from the DP class is the so-called
parity-conserving (PC) class where particles follow a branching
and annihilating random walk with two offsprings (BARW2)
$A\to 3A$, $2A\to \text{\O}$ that conserves the parity of 
their number \cite{Taka92,IJensen93,Avraham,Cardy}.  
Particles following BARW2 may also appear as kinks between 
ordered domains in systems exhibiting two absorbing states  
\cite{Gras84,Men94,Park94,Hin97,Bassler}.
However it was realized that the BARW2 dynamics alone 
is not sufficient condition for a transition
in this class but an exact $Z_2$ symmetry between
the two absorbing states is necessary, too 
\cite{Hin97,Park95,MeOd96,OdMe98,ujMeOd}.

Recently a study on the annihilation fission (AF) process 
$2A\to\text{\O}$, $2A\to 3A$ \cite{Carlon99}
suggested that neither the BARW2 dynamics nor the
$Z_2$ symmetry but simply the occurrence of two absorbing
states can result in a PC class transition.
This model was introduced first by \cite{HT97} 
in the context of a renormalization group analysis of 
the corresponding bosonic field theory.
The main idea was to interpolate between systems with
real and imaginary noise. These studies predicted a non-DP
class transition, but they could not tell to which 
universality class this transition really belongs.

A fermionic version of the AF process introduced in 
\cite{Carlon99} is controlled by two parameters, namely 
the probability of pair annihilation $p$ and the 
probability of particle diffusion $d$. The dynamical 
rules are
\begin{equation}
\begin{split}
AA\text{\O},\,\text{\O} AA \rightarrow AAA  \qquad \text{with rate}
\, & (1-p)(1-d)/2 \\
AA \rightarrow \text{\O\O} \qquad  \text{with rate}\,  &
p(1-d) \\
A\text{\O} \leftrightarrow \text{\O}A \qquad \text{with
rate}\,  & d \ .
\end{split}
\label{DynamicRules}
\end{equation}
Carlon et al. suggested that the phase
diagram can be separated into two regions. 
For low values of $d$ (less than approximately 0.3), 
they found a continuous phase transition
belonging to the PC universality class. 
For large values of $d$, however, a
first order transition was reported.
This claim is based on mean-field, pair mean-field and 
density matrix renormalization group method (DMRG) 
\cite{DMRG} calculations.
An even more recent preliminary numerical study
\cite{Hayepcpd} found non-PC class like critical exponents 
and posed the question if there is a new type of 
non-equilibrium phase transition occurring here.

In this paper I report mored detailed numerical results:
Simulations and generalized mean-field (GMF) approximations 
with coherent anomaly (CAM) extrapolations that give numerical 
evidence of a rich phase diagram with two different kinds 
of new universality classes in the low $d$ and high $d$ regions.

\begin{figure}[h]
\epsfxsize=70mm
\epsffile{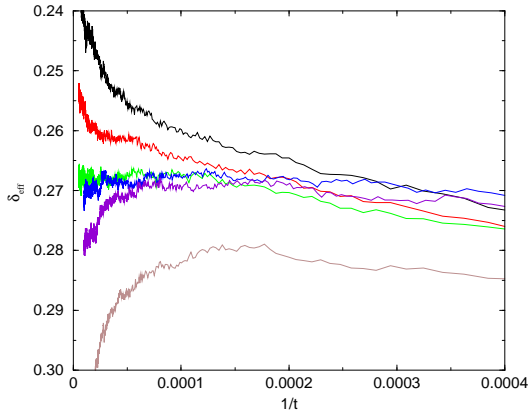}
\vspace{2mm}
\caption{Local slopes of the particle density decay
in the parallel update model at $d=0.2$. Different curves
correspond to $p=0.2481$, $0.24807$, $0.24805$, $0.24802$, 
$0.248$, $0.24795$ (from bottom to top). 
The simulations were performed on a ring of size $L=24000$ 
up to $4\times 10^5$ MCS. Throughout the whole paper 
$t$ is measured in units of Monte-Carlo sweeps.}
\label{pardec2}
\end{figure}

For simulations a parallel update version of the AF process
was used since the model can effectively be mapped on a 
massively parallel processor ring \cite{ASP}. 
To avoid collisions by simultaneous updates the annihilation 
and exchange steps are done on two sub-lattices,
while creation is performed on three sub-lattices 
(with a rate double as in eq.\ref{DynamicRules}).
The critical point has been determined for 
$d=0.05, 0.1, 0.2, 0.5, 0.9$ by following the decay of 
the particles from a uniform, random initial
state ($\rho(t)\propto t^{-\delta}$). 
The local slopes curve of the density
\begin{equation}
\delta_{eff}(t) = {- \ln \left[ \rho(t) / \rho(t/m) \right] 
\over \ln(m)} \label{slopes}
\end{equation}
(where we use $m=8$ usually) at the critical point goes to 
exponent $\delta$ by a straight line, while in 
sub(super)-critical cases it veers down(up) respectively. 
As Fig. \ref{pardec2} shows at $d=0.2$ and $p=0.24802$
this quantity scales without any relevant corrections
(i.e. the straight line is horizontal) and in the 
$1/t\to 0$ limit goes to $\delta=0.268(2)$.
Very similar results has been obtained for $d=0.05$ and $d=0.1$
showing that the exponent $\delta$ is close to the corresponding
value of the PC class ($0.285(10)$ \cite{IJensen93}), 
but appears to be significantly smaller. This $\delta$ 
is in agreement within error margin with the value 
($\sim 0.272(18)$) one can obtain from DMGR calculations 
\cite{Carlon99} using scaling laws 
$\delta=\beta/\nu_{||}=\beta/\nu_{\perp}/z$.

\begin{figure}[h]
\epsfxsize=70mm
\epsffile{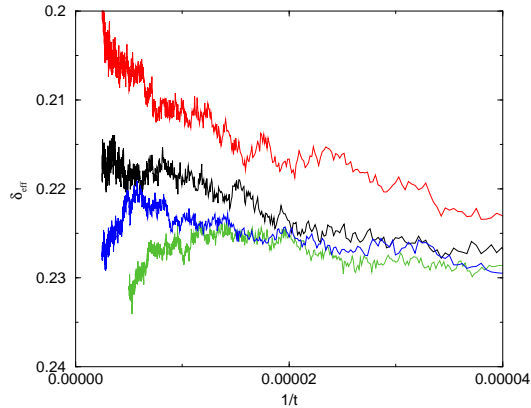}
\vspace{2mm}
\caption{The same as Fig. \ref{pardec2} at $d=0.5$.
Different curves correspond to $p=0.2796$, $0.27958$, $0.27955$, $0.2795$ 
(from bottom to top).}
\label{pardec5}
\end{figure}

For stronger diffusion rates $d=0.5$ and $d=0.9$ one can observe
slower density decays with exponent $\delta=0.215(10)$ 
(Fig. \ref{pardec5}).
Note that I checked, that changing the abscissa on the local
slopes graphs from $1/t$ to $\ln(t)/t$ does not eliminate the
differences between small and large $d$ behaviors. 
Also by assuming strong correction to scaling I could not get
coherent estimates for $\beta$ and $\delta$ that would be 
in agreement with DP or PC classes.
This value of $\delta$ is in agreement with that of the 
coupled DP+annihilation random walk suggested by \cite{HayeDP-ARW}.
This model was proposed to explain the space-time behavior of
the AF with the help of a multi-component model, 
where $A$ particles (corresponding to pairs in AF) perform BARW process:
$A\to 2A$, $A\to \text{\O}$ and occasionally create $B$ particles 
(corresponding to lonely particles in AF)
who follow random walk + annihilation : $2B\to A$ (ARW).
The space-time evolution picture of this model looks similar to that
of AF (and very different from that of BARW2) 
but the numerical simulations 
predicted somewhat different exponents as those of the AF 
\cite{Carlon99,Hayepcpd,Graspcpd,Mendes}.
By approaching $d=1$ corrections to scaling are getting stronger 
and this might leaded to the conclusion of \cite{Carlon99}
that for $d > 0.3$ the transition is of first order.
The present study can not confirm those predictions based on extrapolations
to DMRG results of small system sizes ($L<100$).
The first order transition for $d>0.3$ appears to be unlikely 
also because the mean-field approximation of the model
exhibits continuous phase transition (with $\beta_{MF}=1$) and that 
corresponds to the strong diffusion limit without fluctuations.

\begin{figure}[h]
\epsfxsize=70mm
\epsffile{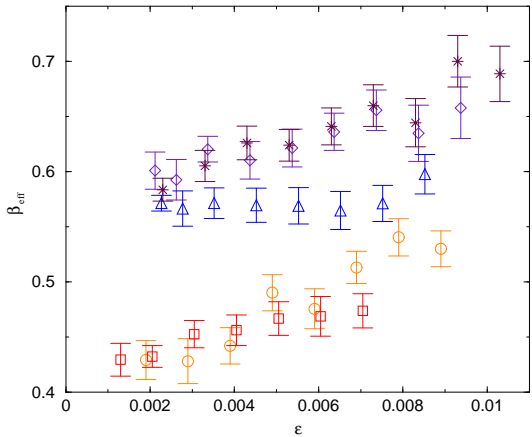}
\vspace{4mm}
\caption{Order parameter exponent results of the AF model for
$d=0.9$ (circles), $d=0.5$ (squares), $d=0.2$ (diamonds),
$d=0.1$ (triangles). $d=0.05$ (stars).
The simulations were performed on a ring of $L=24000$ sites
such that averaging was done following the steady state is
reached for $1000$ samples}
\label{beta}
\end{figure}

The differences from PC class and from first order transition 
can be seen even more clearly by direct measurements of the 
order parameter $\beta$ exponent 
($\rho(\infty)\propto\epsilon^{\beta}$). 
By looking at the effective exponent defined as
\begin{equation}
\beta_{eff}(\epsilon_i) = \frac {\ln \rho(\epsilon_i) -
\ln \rho(\epsilon_{i-1})} {\ln \epsilon_i - \ln \epsilon_{i-1}} \ \ ,
\end{equation}
on Fig. \ref{beta} one can observe two regions 
again: for $d<0.3$ the $\beta_{eff}$ 
tends to $0.58(2)$ as $\epsilon=(p_c-p)\to 0$, while 
for $d>0.3$ it goes to $\sim 0.4(2)$.
 
For small diffusion rates the value $0.58$ is very 
different from that of the PC class ($0.94(10)$) 
\cite{IJensen93,ujMeOd}) and from that of DP class 
($0.2765(1))$\cite{Gras82}, but is in agreement with 
preliminary simulation results of \cite{Hayepcpd} 
and \cite{Mendes}.
For strong diffusion it coincides fairly well with the value
$0.38(6)$ that was found for the coupled DP+ARW model 
\cite{HayeDP-ARW}.

The results for the order parameter exponent has been verified by
generalized mean-field (GMF) approximations \cite{gut87,dic88}
with coherent anomaly extrapolation (CAM) \cite{suz86}.
This method has been proven to give precise estimates for the DP
\cite{GMFDP} and PC \cite{MeOd95} classes. 
The details of the method are described in \cite{ujMeOd} and 
a forthcoming paper \cite{forth} will discuss
the results and show estimates for other exponents as well.

\begin{figure}[h]
\epsfxsize=70mm
\epsffile{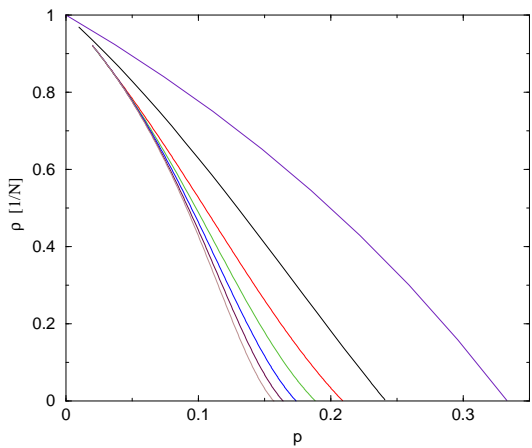}
\vspace{2mm}
\caption{GMF approximations of the particle density
in the symmetrized AF model at $d=0.5$
for $N=1,2, ...7$ (from right to left curves).}
\label{rho_5}
\end{figure}

For technical reasons a variant of the AF model was investigated 
where the creation is symmetric: $A\text{\O}A \to AAA$. This does not
change the essence since the 
$AA\text{\O}/\text{\O}AA \to AAA$ process can be 
decomposed into a $A\text{\O} \to \text{\O}A$ exchange 
and a symmetric creation.
One can easily check that the mean field approximation 
and result of it is the same
\begin{equation}
\rho(\infty)= (1-3 p)/(1-p)
\end{equation}
Higher order approximations for $N=2,3,..7$ converge to the
simulation results of this variant (Fig. \ref{rho_5}) and
the phase diagram qualitatively agrees with that of ref.
\cite{Carlon99}. The leading order singularity at 
critical point remains mean-field type i.e. $\beta_{GMF}=1$.
Furthermore dynamical MC simulation of this variant gives 
the same critical decay behavior as the original AF model.
According to CAM the amplitudes of these singularities $a(N)$
scale in such a way that
\begin{equation}
a(N) \propto (p_c(N)-p_c)^{\beta - \beta_{MF}} \label{anoscal}
\end{equation}
the exponent of true singular behavior can be estimated.
The CAM extrapolation has been applied for the highest 
levels of approximations ($N=4,5,6,7$) and resulted in 
$\beta_{CAM}$ values is agreement with those of simulations 
(see summarizing Table \ref{tab}).
\begin{table}
\caption{Summary of results. The non-universal critical parameter 
$p_c$ of the parallel model is shown here.}
\label{tab}
\begin{tabular}{|l|r|r|r|r|r|}
$d$          &  0.05 & 0.1   & 0.2      &  0.5 & 0.9\\
\hline
$p_c$        &0.25078&0.24889&0.24802&0.27955&0.4324\\ 
$\beta_{CAM}$&  -    &0.58(6)& 0.58(2)  &0.42(4)& -   \\
$\beta$      &0.57(2)&0.58(1)& 0.58(1)  & 0.4(2)&0.39(2)\\
$\delta$     &0.273(2)&0.275(4)&0.268(2)&0.21(1)&0.20(1)\\
$\eta$       &0.10(2) &   -    & 0.14(1)  &0.23(2)&0.48(1)\\
$\delta\prime$&0.004(6)&   -    &0.004(6)  &0.008(9)&0.01(1)\\ 
\end{tabular}
\end{table}
Note, that the CAM extrapolation does not work for $d>0.5$ and
$d<0.1$, because corrections to scaling are getting strong, 
and it is very difficult to find the steady state solutions 
of the GMF equations for $N > 7$.

\begin{figure}[h]
\epsfxsize=70mm
\epsffile{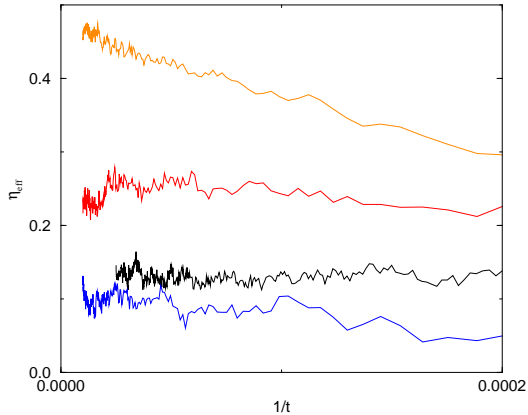}
\vspace{2mm}
\caption{Local slopes of the particle number in seed simulations
in the parallel update model at $d=0.05$, $d=0.2$, $d=0.5$ 
and $d=0.9$ (from bottom to top). 
The simulations were performed on a lattice of size $L=24000$ 
up to $10^5$ MCS.}
\label{eta}
\end{figure}

Time dependent simulations from a single active seed for these models
have been proven to be a very efficient method \cite{GrasTor} since
the slowly spreading clusters do not exceed the allowed system sizes 
hence do not feel finite size effects. Very precise estimates for
the cluster "mass" exponent $N(t)\propto t^{\eta}$ 
and the cluster survival probability exponent 
$P(t)\propto t^{-\delta\prime}$ 
have been obtained in DP and PC classes 
(see references in \cite{Dick-Mar}).
However example in the case of systems with infinitely 
many absorbing states these exponents seem to depend on initial 
conditions \cite{PCP,Mendes-Marques}. 
A rigorous explanation is still missing.
The simulation results for the cluster mass (Fig. \ref{eta})
show strong dependence of $\eta$ on $d$. For $d=0.5$ and
$d=0.9$, where $\beta$ and $\delta$ are roughly the same
it changes from $0.23(2)$ to $0.48(1)$.
The survival probability exponent $\delta\prime$ is roughly
zero, expressing the fact that a single particle survives
with finite probability (see table \ref{tab}).

The continuously changing exponent $\eta$ is in agreement with 
the preliminary observations of \cite{Hayepcpd}, however 
$\beta$ and the $\delta$ seem to be constant in the region:
$0.05\le d \le 0.2$ and $0.5 \le d \le 0.9$, suggesting two
different universality classes. 

The non-universal behavior of the cluster exponent may not 
be so surprising since it was also observed in the $d=0$ limit
of the AF in the pair contact process \cite{PCP,Mendes-Marques}.

In conclusion numerical simulations and analytical GMF+CAM 
calculations suggest two new universality classes in the
AF model along the phase transition line.
For small diffusion single particles cannot escape clusters
where pair annihilation and creation dominates.
For larger diffusions these single particles will wander
among pair creation-annihilation clusters and the coupled
BARW + ARW description of \cite{HayeDP-ARW} will be valid.
These new universal behaviors may sound uneasy following
two decades when the DP hypothesis has been found to describe
almost every continuous transition to absorbing state, but
do not contradict to it at all. 
In the AF model the absorbing state is not singlet, 
there are two (non symmetric) absorbing states  
where the system can evolve with equal probabilities.
One of them is not completely frozen, but a single particle
diffuses in it. 
My results do not support the conclusions of \cite{Carlon99} 
for a PC class transition without BARW2 process and $Z_2$ symmetry 
nor the first order transition for large $d$, but the exponents 
do not contradict to those of \cite{Carlon99} within 
numerical precision.
It is likely that the two exponent ratios determined by 
\cite{Carlon99} are close to those of PC class accidentally.
The cluster exponent $\eta$ behaves non-universally like
in the pair contact process, the $d=0$ limit of the AF.
Deeper understanding of this model that would reveal hidden 
symmetries responsible for the strange critical behavior of AF
would be highly desirable.

\vspace{3mm}
\noindent
{\bf Acknowledgements:}\\

The author would like to thank E. Carlon, P. Grassberger,
M. Henkel, H. Hinrichsen and J. F. F. Mendes for helpful
discussions.
Support from Hungarian research fund OTKA (Nos. T-25286 and 
T-23552) and from B\'olyai (No. BO/00142/99)
is acknowledged. 
The simulations were performed on Aspex's System-V 
parallel processing system (www.aspex.co.uk).


\begin{thebibliography}{999}

\bibitem{TCM} R. Dickman and T. Tom\'e, Phys. Rev. A {\bf 44}, 4833 (1991);
\bibitem{MeOd95} N. ~Menyh\'ard and G. ~\'Odor, 
J. Phys. A {\bf 28}, 4505 (1995).
\bibitem{OSz96} G. \'Odor,  and A. Szolnoki, 
Phys. Rev. E {\bf 53}, 2231 (1996);
\bibitem{MeOd98} N. ~Menyh\'ard, G. ~\'Odor, 
J. Phys. A {\bf 31}, 6771 (1998).
\bibitem{Oerding99} K. Oerding, F. van Wijland, J.-P. Leroy and
H. J. Hilhorst, cond-mat/9910351.

\bibitem{Jan81} H. K. Janssen, Z. Phys. B {\bf 42}, 151 (1981).
\bibitem{Gras82} P. Grassberger, Z. Phys. B {\bf 47}, 365 (1982).
\bibitem{PCP} I. Jensen and R. Dickman, Phys. Rev. E {\bf 48}, 1710 (1993);
I. Jensen, Phys. Rev. Lett. {\bf 70}, 1465 (1993).
\bibitem{Mendes-Marques} J. F. F. Mendes, R. Dickman, M. Henkel and
M. C. Marques, J. Phys. {\bf A 27}, 3019 (1994).

\bibitem{Taka92} H. Takayasu and A. Yu. Tretyakov, Phys.\ Rev.\ Lett. 
{\bf 68}, 3060, (1992).
\bibitem{IJensen93} I.~Jensen, Phys.Rev.E {\bf 50}, 3623 (1994).
\bibitem{Avraham} D. ben Avraham, F. Leyvraz and S. Redner, 
Phys. Rev. E {\bf 50}, 1843 (1994).
\bibitem{Cardy} J. Cardy and U. T\"auber, Phys. Rev. Lett. {\bf 77}, 4780 (1996)

\bibitem{Gras84} P.~Grassberger, F.~Krause and T.~ von der
Twer, J. Phys. A:Math.Gen., L105 {\bf 17} (1984).
\bibitem{Men94} N.~ Menyh\'ard, J.Phys.A:Math.Gen., {\bf 27}, 6139(1994)
\bibitem{Park94} H. M. Kim and H. Park,  Phys.\ Rev.\ Lett. 
{\bf 73}, 2579, (1994).
\bibitem{Hin97} H.~Hinrichsen, Phys. Rev. {\bf E 55}, 219 (1997).
\bibitem{Bassler} K. E. Bassler and D. A. Browne, 
Phys. Rev. Lett. {\bf 77}, 4094 (1996).

\bibitem{Park95} H. Park and H. Park, Physica A {\bf 221}, 97 (1995).
\bibitem{MeOd96} N.~ Menyh\'ard and G. ~\'Odor, J.Phys.A:Math.Gen.
{\bf 29}, 7739 (1996).
\bibitem{OdMe98} G. \'Odor  and N.~Menyh\'ard  Phys. Rev. E {\bf 57} (1998) 5168.
\bibitem{ujMeOd}N.~Menyh\'ard N. and G.~\'Odor, cond-mat/0001101.

\bibitem{Carlon99} E.~Carlon, M.~Henkel, and U.~Schollw{\"o}ck,
\newblock Critical properties of the reaction-diffusion model $2A\rightarrow 3A$, 
$2A\rightarrow$\O, \newblock preprint cond-mat/9912347.

\bibitem{HT97} M.~J. Howard and U.~C. T{\"a}uber,  
{J. Phys.} {\bf A 30}, 7721 (1997).

\bibitem{DMRG} I.~Peschel, X.~Wang, M.~Kaulke, and K.~Hallberg, editors,
\newblock {\em Density-Matrix Renormalization}, Berlin, 1999, Springer.

\bibitem{Hayepcpd} H.~Hinrichsen, \newblock preprint cond-mat/0001177.

\bibitem{ASP} G. \'Odor, A. Krikelis, G. Vesztergombi and F. Rohrbach:
  Proceedings of the 7-th Euromicro workshop on parallel and distributed processing,
  Funchal (Portugal) Feb. 3-5 1999, IEEE Computer society press, Los Alamitos,
  ed.: B. Werner; e-print: physics/9909054

\bibitem{HayeDP-ARW} H.~Hinrichsen, \newblock
Cyclically coupled spreading and pair annihilation, \newblock
private communication.

\bibitem{Graspcpd} P.~Grassberger, private communication.
\bibitem{Mendes} J. F. F. Mendes, private communication.

\bibitem{gut87} H.A. Gutowitz, J.D. Victor and B.W. Knight, 
Physica {\bf 28D}, (1987), 18
\bibitem{dic88} R. Dickman, Phys.Rev. {\bf A38} (1988) 2588.
\bibitem{suz86} M. Suzuki, J.\ Phys.\ Soc. Jpn.\ {\bf 55}, 4205 (1986).
\bibitem{GMFDP} G. \'Odor, Phys. Rev. {\bf E51} (1995) 6261;
G. Szab\'o and G. \'Odor Phys.Rev. {\bf E59} (1994) 2764.

\bibitem{forth} G. \'Odor in preparation.

\bibitem{GrasTor} P.~Grassberger and A.~ de la Torre, Ann.Phys.(NY) {\bf 122}
{\bf 122}, 373 (1979).

\bibitem{Dick-Mar} J.~Marro and R.~Dickman,
\newblock {\em Nonequilibrium phase transitions in lattice models},
\newblock Cambridge University Press, Cambridge, 1999.

\end{thebibliography}
\end{document}